\documentclass[conference]{IEEEtran}
\IEEEoverridecommandlockouts

\usepackage[letterpaper, left=1in, right=1in, bottom=1in, top=0.75in]{geometry}

\usepackage{algorithmic}
\usepackage{amsthm}
\usepackage{amsmath}
\usepackage{amssymb}
\usepackage{amsfonts}
\usepackage{balance}
\usepackage{caption}
\usepackage{cite}
\usepackage{color}
\usepackage{epstopdf}
\usepackage{esint}
\usepackage{graphicx}
\usepackage{subcaption}

\usepackage{epstopdf}

\usepackage{textcomp}
\def\BibTeX{{\rm B\kern-.05em{\sc i\kern-.025em b}\kern-.08em
    T\kern-.1667em\lower.7ex\hbox{E}\kern-.125emX}}
    
\theoremstyle{plain}
\newtheorem{thm}{\protect\theoremname}

\makeatother

\providecommand{\theoremname}{Theorem}

\theoremstyle{plain}
\newtheorem{lem}{\protect\lemmaname}

\makeatother

\providecommand{\lemmaname}{Lemma}

\theoremstyle{plain}

\theoremstyle{plain}

\makeatother

\providecommand{\corname}{Corollary}

\theoremstyle{plain}

\makeatother

\providecommand{\propositionname}{Proposition}

\begin{document}

\title{Outage Probability Analysis of \\ THz Relaying Systems
	\thanks{This work has received funding from the European Commission's Horizon 2020 research and innovation programme under grant agreement No. 761794.}
}
\author{\IEEEauthorblockN{ Alexandros--Apostolos A. Boulogeorgos,  and Angeliki Alexiou}
	\IEEEauthorblockA{\textit{Department of Digital Systems,
			University of Piraeus}, 
		Piraeus, Greece \\
		E-mails: al.boulogeorgos@ieee.org, alexiou@unipi.gr}
}

\maketitle

\begin{abstract}
This paper focuses on quantifying the outage performance of terahertz (THz) relaying systems. In this direction, novel closed-form expressions for the outage probability of a dual-hop relaying systems, in which both the source-relay and relay-destination links suffer from fading and stochastic beam misalignment, are extracted. Our results reveal the importance of taking into account the impact of beam misalignment when characterizing the outage performance of the system as well as when selecting the transmission frequencies.      
\end{abstract}


\section{Introduction}\label{S:Intro}

{D}{espite} the adaptation of several game changing technologies, the envelop of  fifth generation (5G) systems capabilities has been proven to be defined by the available bandwidth~\cite{A:Will_densification_be_the_death_of_5G,WP:Wireless_Thz_system_architecture_for_networks_beyond_5G,A:LC_CR_vs_SS,PhD:Boulogeorgos,Boulogeorgos2020}. Motivated by this, next generation networks are expected to exploit higher frequency bands, such as the terahertz (THz) one~\cite{Boulogeorgos2018,Papasotiriou2020,Kokkoniemi2020,Papsotiriou2019,Boulogeorgos2019a}. However, THz links experience high channel attenuation, which significantly limits the communication range~\cite{our_spawc_paper_2018,PIMRC_275-400,C:UserAssociationInUltraDenseTHzNetworks,Stratidakis2019}. 

To overcome the aforementioned limitation, researchers have recently turned their attention to relaying systems~\cite{Abbasi2017,A:Relay_assisted_nanoscale_communication_in_the_THz_band,P.Monteiro2018,Xia2017,Boulogeorgos2019}. In more detail, in~\cite{Abbasi2017} and~\cite{A:Relay_assisted_nanoscale_communication_in_the_THz_band}, the authors used Monte Carlo simulations in order to quantify the outage performance of a relay-assisted in-vivo nano-scale network in the absence and presence of direct link that operates in the THz band. For the same system model, in~\cite{P.Monteiro2018}, the energy efficiency was evaluated. Additionally, in~\cite{Xia2017}, a number of different relaying strategies for macro-scale THz systems was discussed. Finally, in~\cite{Boulogeorgos2019}, the outage and error performance of a mixed microwave-THz wireless systems were~assessed. 

To the best of the author's knowledge, the theoretical framework for the evaluation of THz wireless relaying systems performance remains vastly unexplored. Motivated by this, this paper studies the outage  performance of such systems. In particular, we extract novel closed-form expressions for the system outage probability (OP), which reveal the joint impact of beam misalignment (BM), and multi-path fading.

\subsubsection*{Notations}
The operator $|\cdot|$  denotes the absolute value, whereas $\exp\left(x\right)$  represents the exponential function.
Additionally,  $\sqrt{x}$ returns the square root of $x$, while, $\min\left(\cdot\right)$ stands for the minimum operator.   Meanwhile, the set of the complex numbers is represented by~$\mathbb{C}$. In addition, $P_r\left(\mathcal{A}\right)$ represents the probability that the event $\mathcal{A}$ is valid.  
Finally, the upper and lower incomplete Gamma functions~\cite[eq. (8.350/2), (8.350/3)]{B:Gra_Ryz_Book} are respectively denoted by $\Gamma\left(\cdot, \cdot\right)$ and $\gamma\left(\cdot, \cdot\right)$, while the Gamma function is represented by $\Gamma\left(\cdot\right)$~\cite[eq. (8.310)]{B:Gra_Ryz_Book}.

\section{System  Model}\label{sec:SM}

\begin{figure}
\centering
\includegraphics[width=1\linewidth,trim=0 0 0 0,clip=false]{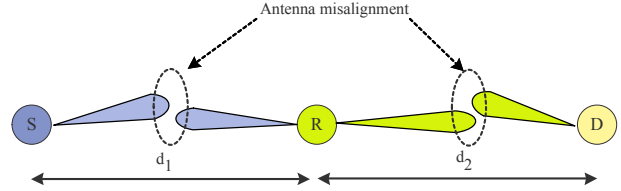}
\caption{System model.}
\label{fig:SM}
\end{figure}

As illustrated in Fig.~\ref{fig:SM}, we consider a dual-hop decode-and-forward (DF) relaying system, in which all nodes operate in the THz band. The source (S), relay (R), and destination (D) nodes are respectively equipped with $N_s$, $N_r$, and $N_d$ antenna elements, and support analog beamforming. Moreover, it is assumed that there is no direct link between S and D. Finally, we assume that R operates in half-duplex mode. As a result, each communication period is divided into two phases. In the first one, S transmits to R, while, in the second one, R decodes, re-encodes and re-transmits the received signal to D. In what follows, for the sake of simplicity, we refer to the S-R and R-D links as the $1-$st and $2-$nd one, respectively.   

During the first phase, the baseband equivalent received signal at R can be expressed~as
\begin{align}
y_r = \mathbf{u}_r^{*} \mathbf{H}_r \mathbf{v}_s s + n_r,
\label{Eq:y_r}
\end{align}
where $n_r$ is the adaptive white Gaussian noise (AWGN) with variance $N_{o,1}$, while $\mathbf{u}_r^{*}$ and $\mathbf{v}_s$ respectively represent the R and S beamforming vectors. Likewise, $s$ stands for the S transmitted signal, while $\mathbf{H}_r$ is the complex MIMO channel~matrix.

Similarly, in the end of the second phase, the  baseband equivalent received signal at R can be obtained~as
\begin{align}
y_d=\mathbf{u}_d^{*} \mathbf{H}_d \mathbf{v}_r \tilde{s} + n_d,
\label{Eq:y_d}
\end{align}
with $n_d$ being the AWGN with variance $N_{o,2}$, whereas $\mathbf{u}_d^{*}$ and $\mathbf{v}_r$ respectively represent the D and R beamforming vectors. Finally, $\tilde{s}$ stand for the re-transmitted by D signal. 

The MIMO channel matrix $\mathbf{H}_i$ with $i\in\{1, 2\}$ can be modeled~as
\begin{align}
\mathbf{H}_i = h_i \mathbf{u}_{i}  \mathbf{v}_i^{*},
\label{Eq:Hi}
\end{align}
where $\mathbf{u}_{1}$ and  $\mathbf{v}_{1}$ are the  $R$ and $S$ perfectly-aligned beamspace directions for the S-R link, while $\mathbf{u}_{2}$ and  $\mathbf{v}_{2}$ stands for the corresponding vectors of the R-D link. Finally, $h_i$ stands for the $i-$th channel coefficient and can be analytically obtained~as
\begin{align}
h_i = 
h_{i}^{(l)} h^{(f)}_{i} h^{(m)}_{i}. 
\label{Eq:h}
\end{align} 

In~\eqref{Eq:h}, $h_{i}^{(l)}$ stands for the deterministic path gain coefficient and can be obtained as~\cite[eqs. (5)-(17)]{A:Analytical_Performance_Assessment_of_THz_Wireless_Systems} 
\begin{align}
h_{i}^{(l)} = \left\{\begin{array}{l l} \frac{c\sqrt{G_{s} G_{r,r}}}{4\pi f_1 d_1} \exp\left(-\frac{1}{2}\beta\left(f_1\right) d_1\right), & \text{ for } i =1  \\ 
\frac{c\sqrt{G_{r,t} G_{d}}}{4\pi f_2 d_2} \exp\left(-\frac{1}{2}\beta\left(f_2\right) d_2\right), & \text{ for } i =2\end{array}\right.,
\end{align}
where $c$ and $\beta$ respectively stand for the speed of light and the molecular absorption coefficient, while $G_{s}$, $G_{r,t}$, $G_{r,r}$, and $G_{d}$ are the source, relay transmission and reception, and destination antenna gains, respectively. Likewise, $f_i$ and $d_i$ with $i\in\{1, 2\}$ are the transmission frequency and distance of the S-R and R-D links. Note that in the $275-425\text{ }\mathrm{GHz}$ range, the molecular absorption coefficient can be accurately evaluated~as~\cite{EuCAP2018_cr_ver7}
\begin{align}
\beta\left(f_i\right) = \sum_{k=1}^{3} u_k\left(f_i\right), \text{ with } i\in\{1, 2\},
\end{align}  
where  
\begin{align}
u_k\left(f_i\right) = \left\{ \begin{array}{l l}\frac{A_k}{B_k + \left(\frac{f_i}{c} - \delta_k \right)}, & \text{for } k\in\{1, 2\}  \\
\sum_{l=0}^3 p_l f^l, & \text{for } k=3\end{array}\right., 
\label{Eq:uk}
\end{align}
with 
\begin{align}
A_k &= \left\{ \begin{array}{l l} g_1 v\left(g_2 v + g_3\right), & \text{for } k=1 \\ 
g_4 v\left(g_5 v + g_6\right),  & \text{for } k=2
\end{array} \right.,
\label{Eq:Ak}
\end{align}
and
\begin{align}
B_k &= \left\{ \begin{array}{l l}\left(g_7 v + g_8\right)^2, & \text{for } k=1 \\ \left(g_9 v + g_{10}\right)^2, & \text{for } k=2 \end{array} \right.
\label{Eq:Bk}
\end{align}
In~\eqref{Eq:uk}, $\delta_1=10.835\text{ }\mathrm{cm}^{-1}$, $\delta_2=12.664\text{ }\mathrm{cm}^{-1}$, $p_0=-6.36\times 10^{-3}$, $p_1=9.06\times 10^{-14}\text{ }\mathrm{s}$, $p_2=-3.94\times 10^{-25}\text{ }\mathrm{s}^2$, $p_3=5.54\times 10^{-37}\text{ }\mathrm{s}^3$. Meanwhile, in~\eqref{Eq:Ak} and~\eqref{Eq:Bk}, $g_1=0.2205$, $g_2=0.1303$, $g_3=0.0294$, $g_4=2.014$, $g_5=0.1702$, $g_6=0.0303$, $g_7=0.4093$, $g_8=0.0925$, $g_9=0.537$, and $g_{10}=0.0956$. Finally, $v$ represents the volume of mixing ratio of the water vapor and can be obtained~as
$v = \phi \frac{p_s\left(T, p_h\right)}{p_h},$
where $p_s$ and $p_h$ are respectively the relative humidity and the atmospheric pressure in $\mathrm{hPa}$. Finally, $p_s\left(T, p_h\right)$ represents the saturated water vapor partial pressure in temperature $T$ and atmospheric pressure $p_h$, and can be expressed~as
$p_s\left(T,p_h\right) = \kappa_1\left(\kappa_2 + \kappa_3 p_h\right) \exp\left(\kappa_4\frac{T-\kappa_5}{T-\kappa_6}\right),$
where $\kappa_1=6.1121\text{ } \mathrm{hPa}$, $\kappa_2=1.0007$, $\kappa_3=3.46\times 10^{-6}\text{ }\mathrm{hPa}^{-1}$, $\kappa_4=17.502$, $\kappa_5=273.15\,^{o}K$, and~$\kappa_6=32.18\,^{o}K$.

 Moreover, $h^{f}_{i}$ represents the fading coefficient. Note that the envelop of $h^{(f)}_i$ is assumed to follow a generalized Gamma distribution with probability density function (PDF) and cumulative density function (CDF) that can be respectively expressed~as  
\begin{align}
f_{\left|h^{(f)}_{i}\right|}(x)=\frac{\alpha_{i} \mu_{i}^{\mu_{i}} x^{\alpha_i \mu_i - 1}}{\left(\hat{h}^{(f)}_{i}\right)^{\alpha_i \mu_i} \Gamma(\mu_i)}  \exp\left(-\mu_i \dfrac{x^{\alpha_i}}{\left(\hat{h}^{(f)}_{i}\right)^{\alpha_i}}\right)
\label{Eq:f_hf1}
\end{align}
and
\begin{align}
F_{\left|h^{(f)}_i\right|}(x) = 1-\frac{\Gamma\left(\mu_i,\mu_i\frac{x^{\alpha_i}}{\left(\hat{h}^{(f)}_{i}\right)^{\alpha_i}}\right)}{\Gamma(\mu_i)},
\label{Eq:F_hf1_definition}
\end{align}
where $\alpha_i > 0$  is a fading parameter, $\mu_i$ is the normalized variance of the fading channel envelope, and $\hat{h}^{(f)}_{i}$ is the $\alpha_i$-root mean value of the fading channel envelop.
Additionally, $h^{(m)}_{i}$ stands for the misalignment fading coefficient and its PDF can be obtained~as
$f_{h^{(m)}_i}(x) = \frac{\zeta_i}{S_{o,i}^{\zeta_i}} x^{\zeta_i-1},\quad 0\leq x \leq S_{o,i},$
where $S_{o,i}$ is the portion of the collected signal at the receiver of the $i-$th link under perfect alignment conditions, whereas $\zeta = \frac{w_i^2}{4\sigma_{s,i}^2}$ with $w_i$ and $\sigma_{s,i}$ respectively being the equivalent beam-width and jitter standard deviation at the receiver plane of the $i-$th link.

Finally, by assuming that, during the initial access phase, the optimal beamforming vectors have been selected; hence, the beam directions are orthogonal, i.e., $\mathbf{u}_{r,1}^{*} \mathbf{u}_1 \mathbf{v}_1^{*} \mathbf{v}_{s,1}=1$, and $\mathbf{u}_{d,2}^{*} \mathbf{u}_2 \mathbf{v}_2^{*} \mathbf{v}_{r,2}=1$, and that the BM can be fully modeled by $h_{i}^{(m)}$, $i\in\{1,2\}$, with the aid of~\eqref{Eq:Hi},~\eqref{Eq:y_r} and~\eqref{Eq:y_d} can be respectively  simplified~as
\begin{align}
y_r = h_1 s + n_r
\end{align} 
and
\begin{align}
y_d = h_2 \tilde{s} + n_d.
\end{align}

\section{Performance analysis}


\subsection{End-to-end SNR statistics} 

Since R operates in DF mode, the equivalent end-to-end SNR can be obtained~as
\begin{align}
\rho_{e} = \min\left(\rho_{1},\rho_{2}\right),
\label{Eq:rho_e}
\end{align}
where $\rho_1$ and $\rho_2$ are respectively the S-R and R-D link SNRs and can be obtained~as 
\begin{align}
\rho_{i} =  \overline{\rho}_i |h^{(fm)}_i|^2, 
\label{Eq:gamma1}
\end{align}
with $i\in\{1, 2\}$, $|h^{(fm)}_i|^2=|h^{(f)}_i|^2 |h^{(m)}_i|^2$, 
and $P_1$, $P_2$ respectively being the S and R transmitted~powers.

The following theorem and lemmas return the CDF of $\rho_e$ for the general  case in which both S-R and R-D links experience BM, as well as for the special cases in which a single or no link suffers for BM. 

\begin{thm}
For the general case, in which both links suffer from BM, the CDF of $\rho_e$ can be evaluated as
\begin{align}
F_{\rho_e}(x) = 
1- \prod_{i=1}^2 \sum_{k=0}^{\mu_i-1} 
&
 \frac{x^{\zeta_i/2}}{\alpha_i {\overline{\rho}_i^{\zeta_i/2} }\left(\hat{h}^{(f)}_{i}\right)^{\zeta_i}}  
\frac{\zeta_i}{S_{o,i}^{\zeta_i}}  
 \frac{\mu_i^{\frac{\zeta_i}{\alpha_i}}}{k!} 
\nonumber \\ &\hspace{-0.7cm}\times 
\Gamma\left(\frac{\alpha_i k-\zeta_i}{\alpha_i},{\frac{ \mu_i S_{o,i}^{-\alpha_i} x^{\alpha_i/2}}{\overline{\rho}_i^{\alpha_i/2} \left(\hat{h}^{(f)}_i\right)^{\alpha_i}}}  \right). 
\label{Eq:F_rho_e_final}
\end{align}
\end{thm}
\begin{IEEEproof}
Please refer to Appendix A. 
\end{IEEEproof}
\begin{lem}
	For the special case in which only one of the S-R and R-D links suffer from BM, the CDF of the end-to-end equivalent SNR can be expressed as~\eqref{Eq:F_rho_e_sc_final}, given at the top of the next page. 
	\begin{figure*}
		\begin{align}
		F_{\rho_e}^{sc}(x)= 1- \sum_{k=0}^{\mu_i-1} 
		&
		\frac{x^{\zeta_i/2}}{\alpha_i {\overline{\rho}_i^{\zeta_i/2} }\left(\hat{h}^{(f)}_{i}\right)^{\zeta_i}}  
		\frac{\zeta_i}{S_{o,i}^{\zeta_i}}  
		\frac{\mu_i^{\frac{\zeta_i}{\alpha_i}}}{k!} 
		\Gamma\left(\frac{\alpha_i k-\zeta_i}{\alpha_i},{\frac{ \mu_i S_{o,i}^{-\alpha_i} x^{\alpha_i/2}}{\overline{\rho}_i^{\alpha_i/2} \left(\hat{h}^{(f)}_i\right)^{\alpha_i}}}  \right)
		\frac{1}{{\Gamma(\mu_j)}}{\Gamma\left(\mu_j,\mu_j\frac{x^{\alpha_j/2}}{\left(\hat{h}^{(f)}_{i}\right)^{\alpha_i} \overline{\rho}_j^{\alpha_j/2}}\right)}
		\label{Eq:F_rho_e_sc_final}
		\end{align}
		\hrulefill
	\end{figure*}
\end{lem}
\begin{IEEEproof}
	Please refer to Appendix B. 
\end{IEEEproof}

\begin{lem}
	For the case in which no-link experience BM, the end-to-end equivalent SNR can be written~as
	\begin{align}
	F_{\rho_{j}^{\text{id}}}(x) = 1 - \prod_{j=1}^{2} \frac{\Gamma\left(\mu_j,\mu_j\frac{x^{\alpha_j/2}}{\left(\hat{h}^{(f)}_{i}\right)^{\alpha_i} \overline{\rho}_j^{\alpha_j/2}}\right)}{{\Gamma(\mu_j)}}.
	\label{Eq:F_rho_e_wm_final}
	\end{align}
\end{lem}
\begin{IEEEproof}
	Please refer to Appendix C. 
\end{IEEEproof}

\subsection{Outage Analysis}\label{sec:PA}

The OP of the equivalent e2e link can be defined as the probability that the equivalent e2e SNR falls below a predetermined threshold, $\rho_{th}$, i.e.
$P_{o}=P_{r}\left(\rho_e \leq \rho_{th}\right),$
or equivalently
\begin{align}
P_{o}= \left\{\begin{array}{l l} 
F_{\rho_e}(\rho_{th}), & \text{ both links suffer from BM}\\
F_{\rho_e}^{sc}(\rho_{th}) & \text{ only one link suffers from BM} \\
F_{\rho_{j}^{\text{id}}}(\rho_{th}) & \text{no link suffers from BM}
\end{array}\right..
\end{align}

\section{Results \& Discussion}

Respective numerical results for different scenarios are presented in this section to highlight the joint effect of antenna misalignment and fading. In more detail, the following insightful scenario is considered. The THz relaying system is operating under standard environmental conditions, i.e. the relative humidity is $50\%$, the atmospheric pressure equals $101325\text{ }\mathrm{Pa}$ and the temperature is set to $296^oK$. Moreover, the operation frequency is $275\text{ }\mathrm{GHz}$, the transmit and receive antenna gains of both the S and R are set to $55\text{ }\mathrm{dBi}$. The S-R and R-D links transmission distances are set to $10\text{ }\mathrm{m}$. Finally, unless otherwise stated, $\alpha_1=\alpha_2=1$, and $\mu_1=\mu_2=3$.  It should be noted that each of the following figures contains both analytical and Monte Carlo results represented by lines and discrete marks,~respectively.


\begin{figure*}
	\centering
	\begin{subfigure}[b]{0.32\textwidth}
		\centering
		\includegraphics[width=1\linewidth,trim=0 0 0 0,clip=false]{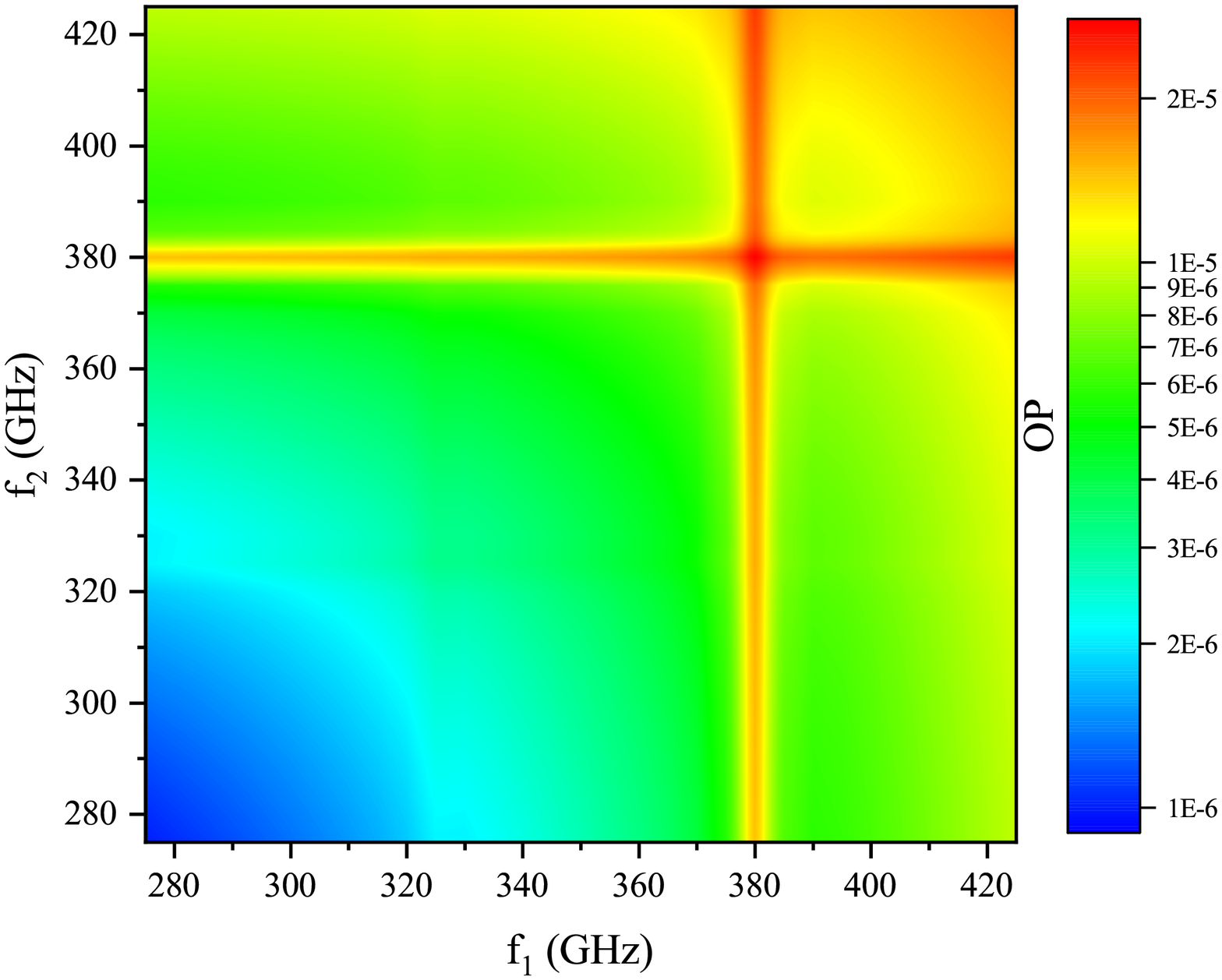}
		\caption{ }
	\end{subfigure}
	\begin{subfigure}[b]{0.32\textwidth}
		\centering
		\includegraphics[width=1\linewidth,trim=0 0 0 0,clip=false]{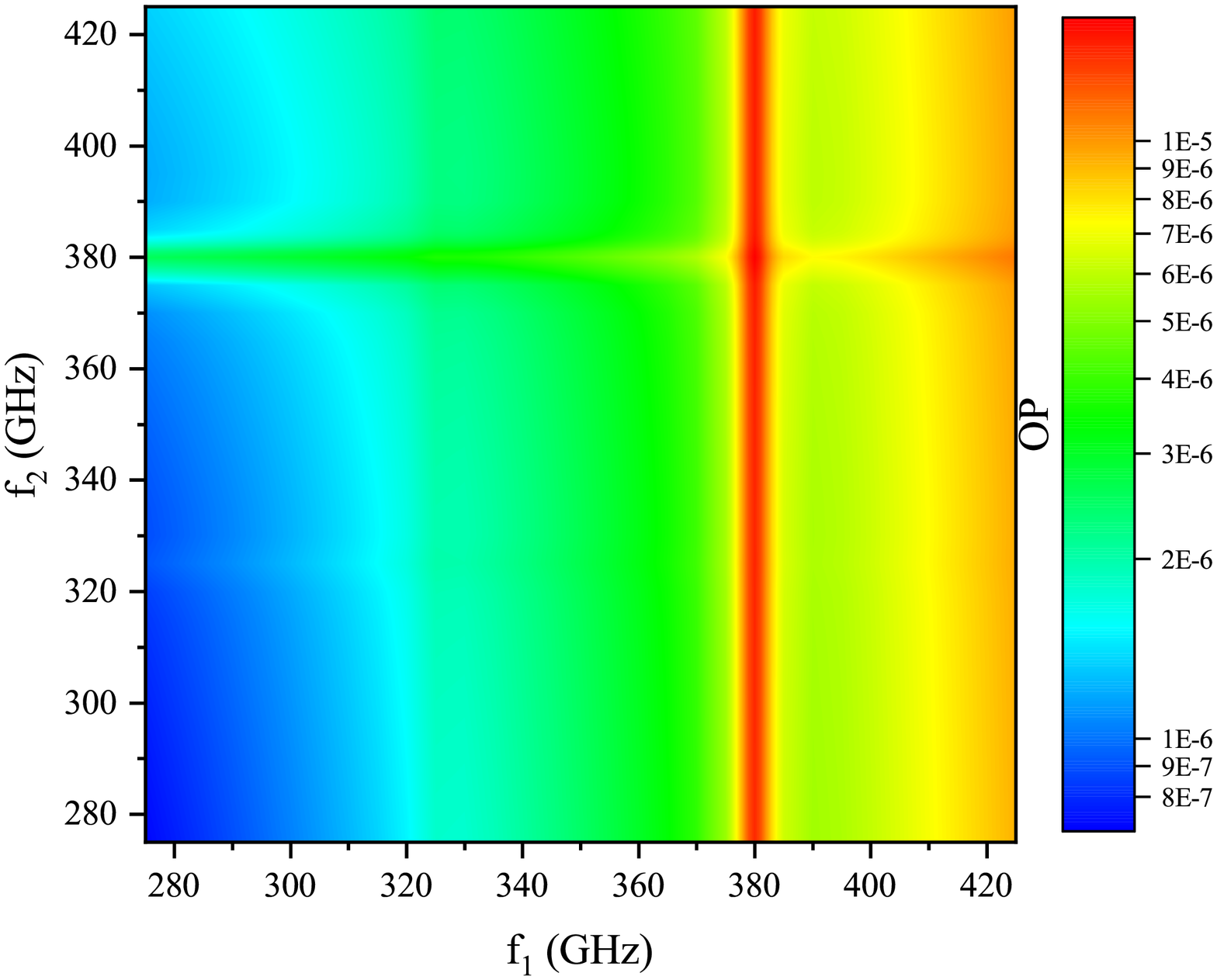}
		\caption{ }
	\end{subfigure}
	\begin{subfigure}[b]{0.32\textwidth}
		\centering
		\includegraphics[width=1\linewidth,trim=0 0 0 0,clip=false]{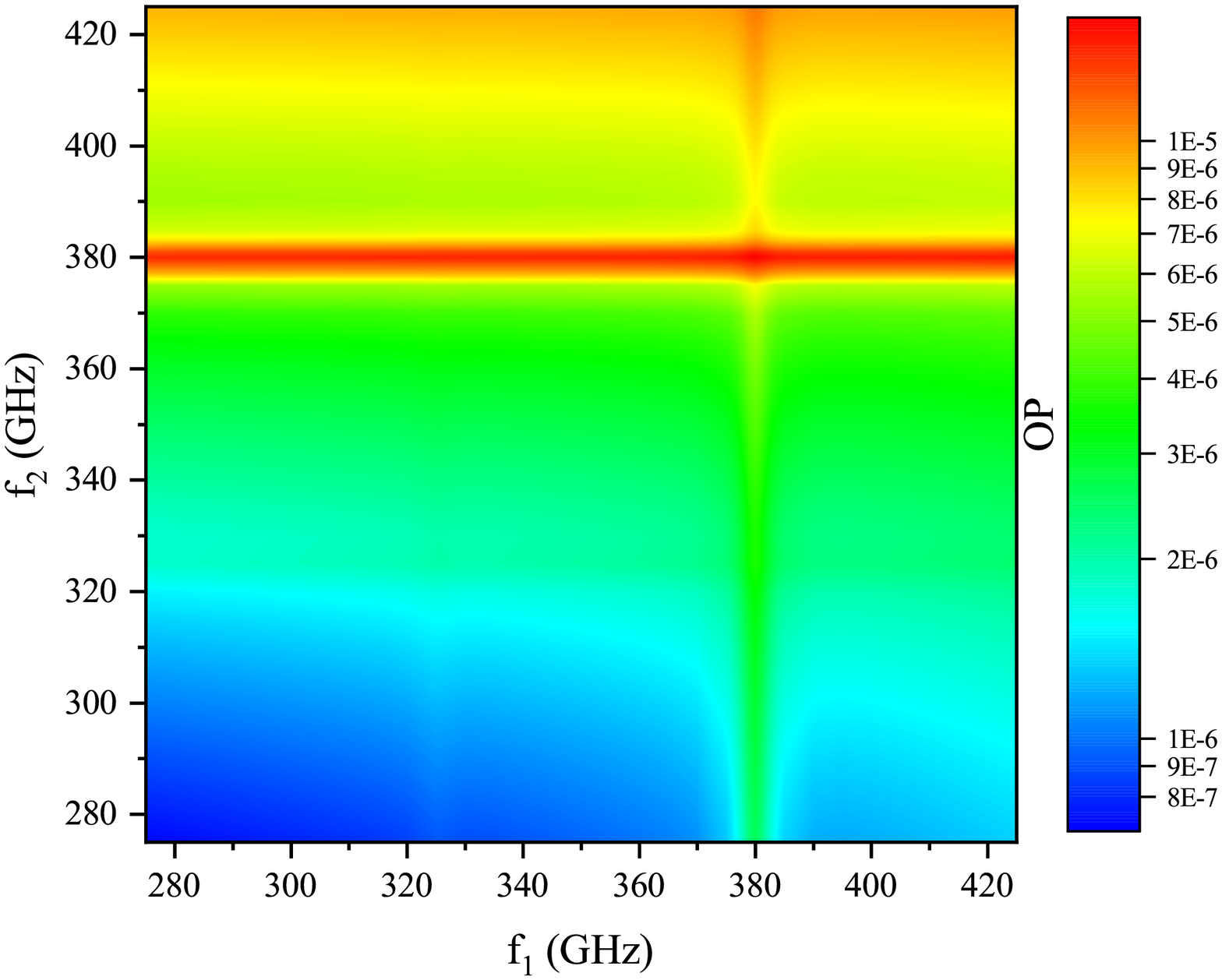}
		\caption{ }
	\end{subfigure}
	\begin{subfigure}[b]{0.32\textwidth}
		\centering
		\includegraphics[width=1\linewidth,trim=0 0 0 0,clip=false]{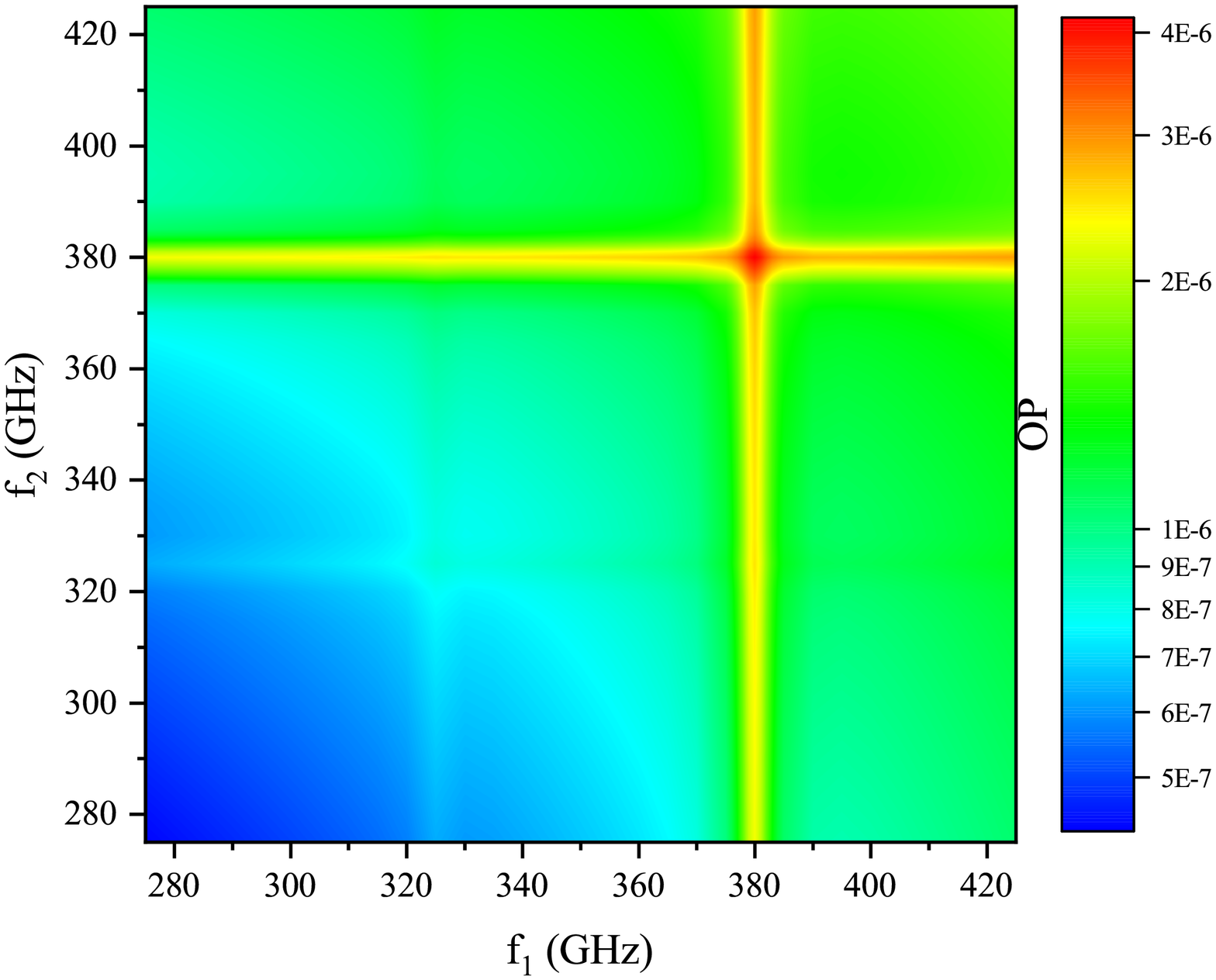}
		\caption{ }
	\end{subfigure}
	\begin{subfigure}[b]{0.32\textwidth}
		\centering
		\includegraphics[width=0.9\linewidth,trim=0 0 0 0,clip=false]{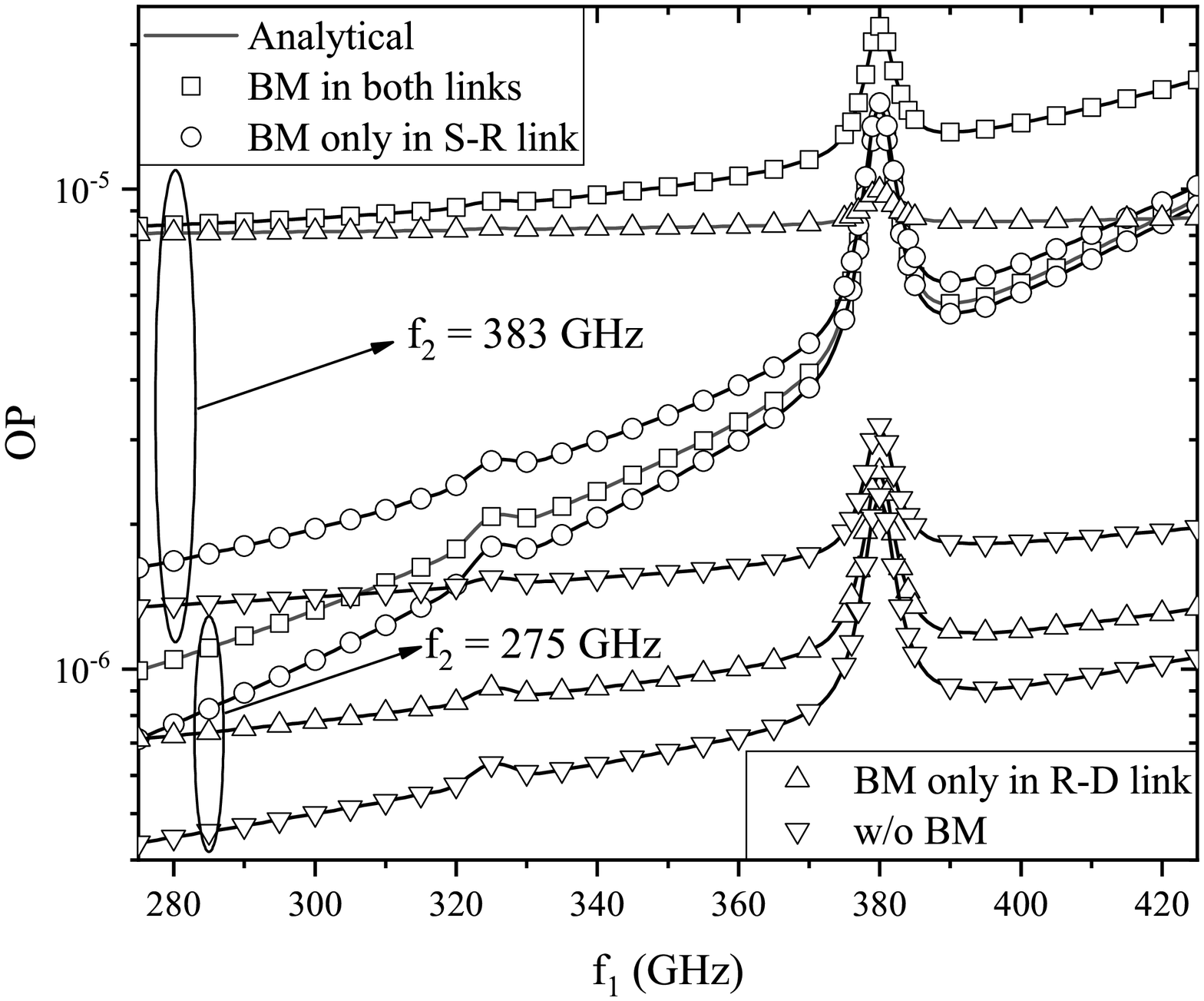}
		\caption{ }
	\end{subfigure}
	\caption{OP vs $f_1$ and $f_2$, for different the following  BM scenarios: (a) both S-R and R-D links experience BM; (b) only S-R link suffers from BM, while R-D is BM-free; (c) only R-D link suffers from BM, while S-R is BM-free; and (d) both S-R and R-D are BM-free. Also, OP vs $f_1$ for different values od $f_2$ and all the aforementioned scenarios (e).}
	\label{fig:OP_vs_f1_and_f2}
	\hrulefill
\end{figure*}

Figure~\ref{fig:OP_vs_f1_and_f2} demonstrates the outage performance of the relaying systems for different values of $f_1$, $f_2$ and BM scenarios. In more detail, in Fig.~\ref{fig:OP_vs_f1_and_f2}.a, the OP is plotted against $f_1$ and $f_2$, assuming that both the S-R and R-D links experience the same levels of BM, i.e. $\sigma_{s,1}=\sigma_{s,2}=1\text{ }\mathrm{cm}$. Similarly, the outage performance for the case in which only the S-R link experience BM of $\sigma_{s,1}=1\text{ }\mathrm{cm}$, while the R-D one is BM-free is plotted in Fig.~\ref{fig:OP_vs_f1_and_f2}.b, whereas the case in which the R-D link suffers from BM of $\sigma_{s,2}=1\text{ }\mathrm{cm}$ and the S-R is BM-free is provided in Fig.~\ref{fig:OP_vs_f1_and_f2}.c. As a benchmark, in~Fig.~\ref{fig:OP_vs_f1_and_f2}.d, the OP performance of the BM-free relaying system are depicted. For the shake of comparison,  in~Fig.~\ref{fig:OP_vs_f1_and_f2}.e, the OP is plotted as a function of $f_1$, for $f_2=275\text{ }\mathrm{GHz}$ and $383\text{ }\mathrm{GHz}$, which are respectively the best and worst case scenario, and for all the aforementioned BM scenarios. From Fig.~\ref{fig:OP_vs_f1_and_f2}.a, it becomes evident that, for the case in which both links suffer from BM, there exist a low OP transmission windows for $f_1,f_2\in[275, 320\text{ }\mathrm{GHz}]$. Moreover, the worst outage performance are met for $f_1=f_2=383\text{ }\mathrm{GHz}$. From Figs.~\ref{fig:OP_vs_f1_and_f2}.b and \ref{fig:OP_vs_f1_and_f2}.c, it is observed that, for a fixed OP requirement, the link that suffers from BM determines the bandwidth of the transmission signal. Moreover, from Fig.~\ref{fig:OP_vs_f1_and_f2}.d, it is revealed that even in the absence of BM, the best and worst outage performance are respectively met for $f_1=f_2=275\text{ }\mathrm{GHz}$ and $f_1=f_2=383 \text{ }\mathrm{GHz}$, where the minimum and maximum path-loss is observed. This indicates the detrimental impact of path-loss on the system performance. Likewise, from Fig.~\ref{fig:OP_vs_f1_and_f2}.e, it becomes evident that, for a given $f_1$, as $f_2$ increases, the OP also increases. Also, for a fixed $f_2$, as $f_1$ increases, the outage performance degrades. Finally, from this figure, it becomes apparent that the joint impact of BM and path-gain should be taken into account when characterizing the outage performance of the THz relaying system.  

\begin{figure}
	\centering
	\includegraphics[width=0.7\linewidth,trim=0 0 0 0,clip=false]{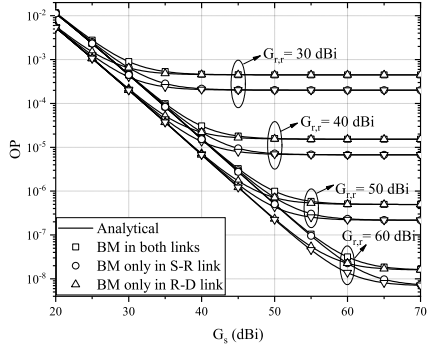}
	\caption{OP vs $G_s$, for different values of $G_{r,r}$ and  BM~scenarios.}
	\label{fig:OP_vs_Gs}
\end{figure} 

Figure~\ref{fig:OP_vs_Gs} illustrates the OP as a function of $G_s$ for different values of $G_{r,r}$ and  BM scenarios, assuming $P_1/N_0/\rho_{th}=P_2/N_0/\rho_{th}=50\text{ }\mathrm{dB}$, $G_{r,t}=50\text{ }\mathrm{dBi}$, and $\sigma_{s,1}=\sigma_{s,2}=1\text{ }\mathrm{cm}$. As expected, for given $G_{r,r}$ and BM scenario, as~$G_{s}$ increases, the deterministic path-gain increases and as a consequence the OP decreases. For example, for the case in which both links suffer from BM and $G_{r,r} = 40\text{ }\mathrm{dBi}$, as $G_{s}$ shifts from $30$ to $50\text{ }\mathrm{dBi}$, the OP decreases for about $30$ times. Similarly, we observe that, for fixed $G_s$ and BM scenario, as $G_{r,r}$ increases, the OP decreases. For instance, for the scenario in which both links experience BM and $G_{s} = 40\text{ }\mathrm{dBi}$, as $G_{r,r}$ moves from $30$ to $50\text{ }\mathrm{dBi}$, the OP decreases approximately $30$ times. By comparing, the aforementioned examples, the reciprocity of the link is revealed. Additionally, from this figure, we observe that there exist a floor on the OP in respect to~$G_s$.        

\begin{figure}
	\centering
	\includegraphics[width=0.7\linewidth,trim=0 0 0 0,clip=false]{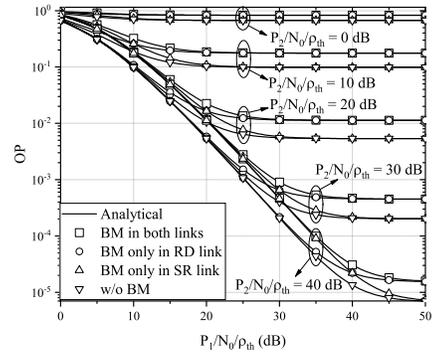}
	\caption{OP vs $P_1/N_0/\rho_{th}$, for different values of $P_{2}/N_0/\rho_{th}$ and  BM scenarios.}
	\label{fig:OP_vs_P1_No_gth}
\end{figure}

Figure~\ref{fig:OP_vs_P1_No_gth} depicts the OP as a function of $P_1/N_0/\rho_{th}$ for different values of $P_2/N_0/\rho_{th}$ and BM scenarios, assuming that $\sigma_{s,1}=\sigma_{s,2}=1\text{ }\mathrm{cm}$. 
In more detail, the following BM scenarios are examined: (i) both links suffer from the same level of BM, (ii) only the SR link experience BM, while RD is BM-free, and (iii) only the R-D link experience BM, while S-R is BM-free. Finally, as a benchmark, the OP for the ideal case, in which both the S-R and R-D links are BM-free, is delivered. From this figure, we observe that, for a fixed $P_2/N_0/\rho_{th}$ and BM scenario, as $P_{1}/N_0/\rho_{th}$ increases, the OP decreases. For example, for the case in which both links suffer for BM and $P_{2}/N_0/\rho_{th}=30\text{ }\mathrm{dB}$, as $P_{1}/N_0/\rho_{th}$ increases from $20$ to $30\text{ }\mathrm{dB}$, the outage performance improves for about $10$ times. Moreover, it becomes apparent that for $P_{1}/N_0/\rho_{th}\geq P_{2}/N_0/\rho_{th}$, there exists an OP floor that depends on the quality of the RD link. As a result, in the $P_{1}/N_0/\rho_{th}\geq P_{2}/N_0/\rho_{th}$ regime, systems in which only the R-D link suffers from BM, and the corresponding ones that experience BM in both the S-R and R-D links have approximately the same performance,  whereas, the same occurs for systems in which only the S-R link suffers from BM and systems that both links are BM-free. Additionally, for a given $P_{1}/N_0/\rho_{th}$, as $P_{2}/N_0/\rho_{th}$ increases, the outage performance improves. For instance, for the case in which both links suffer from BM and $P_{1}/N_0/\rho_{th} = 20\text{ }\mathrm{dB}$, a $P_{2}/N_0/\rho_{th}$ increase from $10$ to $20\text{ }\mathrm{dB}$ results at more than $10$ times OP decrease. A similar outage performance improvement also occurs for the cases in which only one or no-link suffers from BM, under the same $P_{2}/N_0/\rho_{th}$ shift. Finally, we observe that there exist an important performance difference between THz relaying systems that suffer from BM and the corresponding ideal ones. For example, for $P_{1}/N_0/\rho_{th}=35\text{ }\mathrm{dB}$ and $P_{2}/N_0/\rho_{th}=30\text{ }\mathrm{dB}$, the system in which only the R-D link suffers from BM achieves $9\%$ better outage performance in comparison with the one that both links experience BM and about $50\%$ worst performance than the one that BM exists in the SR link. This indicates the importance of accurately modeling BM in THz relaying systems. 

\begin{figure}
	\centering
	\includegraphics[width=0.7\linewidth,trim=0 0 0 0,clip=false]{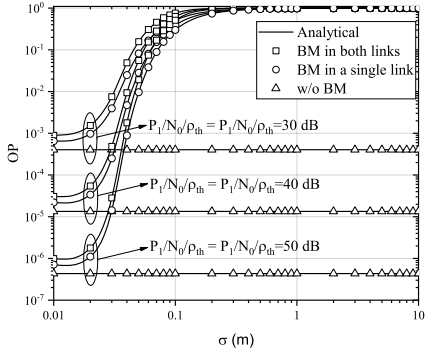}
	\caption{OP vs $\sigma$, for different values of $P_{1}/N_0/\rho_{th}=P_{2}/N_0/\rho_{th}$ and BM scenarios.}
	\label{fig:OP_vs_sigma}
\end{figure} 

Figure~\ref{fig:OP_vs_sigma} illustrates the OP as a function of $\sigma_{s,1}=\sigma_{s,2}=\sigma$, for different values of $P_{1}/N_0/\rho_{th}=P_{2}/N_0/\rho_{th}$ and the following BM scenarios: (i) both the S-R and R-D links suffer from the same level of BM, (ii) either the S-R or the R-D link experience BM, and (iii) both links are BM-free. As expected, for a given $P_{1}/N_0/\rho_{th}=P_{2}/N_0/\rho_{th}$, as $\sigma$ increases, the OP also increases. For example, for $P_{1}/N_0/\rho_{th}=P_{2}/N_0/\rho_{th}=50\text{ }\mathrm{dB}$ and the case in which both links suffer from BM, as $\sigma$ increases from $1$ to $5\text{ }\mathrm{cm}$, the OP also increases from $9.9\times 10^{-7}$ to $1.9\times 10^{-2}$. Additionally, for a given $\sigma$ and BM scenario, as $P_{1}/N_0/\rho_{th}=P_{2}/N_0/\rho_{th}$ increases, the outage performance improves. For instance, for $\sigma=5\text{ }\mathrm{cm}$, in the case in which both links experience BM, the OP decreases from $5.6\times 10^{-2}$ to $1.9\times 10^{-2}$, as $P_{1}/N_0/\rho_{th}=P_{2}/N_0/\rho_{th}$ shifts from $40$ to $50\text{ }\mathrm{dB}$. Similarly, for the case in which only one link suffers from BM, the same $\sigma$ and $P_{1}/N_0/\rho_{th}=P_{2}/N_0/\rho_{th}$ increase results to an OP increase from $2.8\times 10^{-2}$ to $9.69\times 10^{-3}$. Likewise, from this figure, it becomes evident that THz relaying systems in which only one link suffers from BM outperforms the ones that both links experience BM, for the same levels of BM and $P_{1}/N_0/\rho_{th}=P_{2}/N_0/\rho_{th}$. Finally, the BM-free THz relaying system outperforms both the aforementioned~scenarios.              

\begin{figure}
	\centering
	\includegraphics[width=0.85\linewidth,trim=0 0 0 0,clip=false]{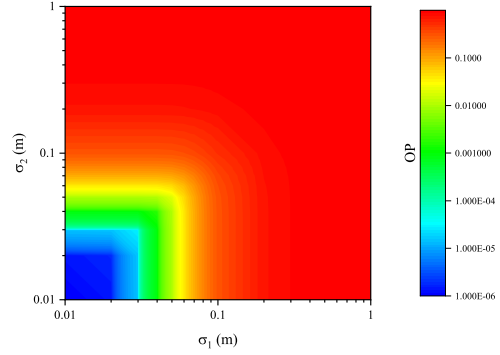}
	\caption{OP vs $\sigma_1$ and $\sigma_2$.}
	\label{fig:OP_vs_sigma1_and_sigma2}
\end{figure}

Figure~\ref{fig:OP_vs_sigma1_and_sigma2} plots the OP as a function of $\sigma_1$ and $\sigma_2$, assuming  $P_{1}/N_0/\rho_{th}=P_{2}/N_0/\rho_{th}=50\text{ }\mathrm{dB}$. As expected, for a fixed $\sigma_1$, as $\sigma_2$ increases, the quality of the R-D link degrades; hence, the OP increases. For instance, for $\sigma_1=5\text{ }\mathrm{cm}$, as $\sigma_2$ shifts from $1$ to $5\text{ }\mathrm{cm}$, the OP changes from $9.69\times 10^{-3}$ to $1.93\times 10^{-2}$. Similarly, for a fixed $\sigma_2$, as $\sigma_1$ increases, an outage performance degradation occurs. For example, for $\sigma_2=2\text{ }\mathrm{cm}$, as $\sigma_1$ increases from $1$ to $5\text{ }\mathrm{cm}$, the OP also increases for about ten times.   

\begin{figure}
	\centering
	\includegraphics[width=0.7\linewidth,trim=0 0 0 0,clip=false]{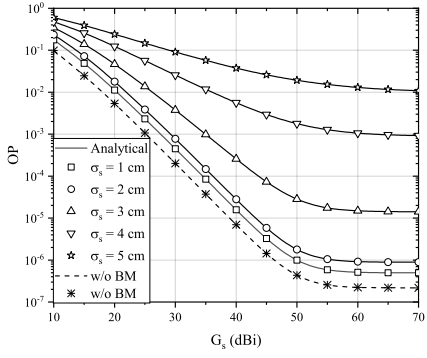}
	\caption{OP vs $G_s$, for different values of $\sigma_s$.}
	\label{fig:OP_vs_Gs_ss}
\end{figure} 

In Fig.~\ref{fig:OP_vs_Gs_ss}, the OP is plotted as a function of $G_s$, for different values of $\sigma_s$, assuming  $P_{1}/N_0/\rho_{th}=P_{2}/N_0/\rho_{th}=50\text{ }\mathrm{dB}$, $\sigma_{s,1}=\sigma_{s,2}=\sigma_s$ and that both links suffer from BM. As a benchmark, the case in which both links are BM-free is also plotted. As expected, for a given $\sigma_s$, as~$G_{s}$ increases, the OP decreases. Moreover, for a fixed $G_{s}$, as $\sigma_s$ increases, the outage performance degrades. Finally, an OP floor in the high $G_s$ regime is observed.  

\section{Conclusions}

In this paper, we presented novel closed-form expressions for the OP of THz wireless relaying systems, which take into account the intermediate links particularities, namely deterministic path-loss, stochastic BM, and fading. Our results highlighted the importance of accurate characterization of the intermediate channels, when assessing and designing such systems.  

\section*{Appendices}
\section{Appendix A}

In general, the CDF of $\rho_e$ can be evaluated~as
$F_{\rho_e}(x) = \Pr\left(\rho_e \leq x\right),$
which, by employing~\eqref{Eq:rho_e}, can be rewritten as
\begin{align}
F_{\rho_e}(x) = \Pr\left(\min\left(\rho_1, \rho_2\right) \leq x \right),
\label{Eq:F_rho_e_1} 
\end{align}
or equivalently
\begin{align}
F_{\rho_e}(x) = 1 - \Pr\left(\min\left(\rho_1, \rho_2\right) \geq x \right). 
\label{Eq:Frho_e_1_s2}
\end{align}
The event~$\mathcal{U}_1 = \left\{\min\left(\rho_1, \rho_2\right)\geq x\right\}$ can be rewritten as $\mathcal{U}_1 =\left\{ \rho_1 \geq x \right\}\cap\left\{ \rho_2 \geq x \right\}$, Moreover, since $|h_1|$ and $|h_2|$ are independent random variables (RVs), $\rho_1$ and $\rho_2$ are also independent RVs; hence,~\eqref{Eq:Frho_e_1_s2} can be simplified~as
 \begin{align}
F_{\rho_e}(x) = 1 - \Pr\left(\left\{ \rho_1 \geq x \right\} \right) \Pr\left(\left\{ \rho_2 \geq x \right\} \right), 
\label{Eq:Fgamma_e3}
\end{align}
or equivalently
\begin{align}
F_{\rho_e}(x) = 1 - \left(1-\Pr\left(\left\{ \rho_1 \leq x \right\} \right)\right) \left(1- \Pr\left(\left\{ \rho_2 \leq x \right\} \right)\right). 
\label{Eq:Frho_e4}
\end{align}
By taking into account that $\Pr\left(\left\{ \rho_i \leq x \right\} \right)=F_{\rho_i}(x)$ with $i\in\{1, 2\}$,~\eqref{Eq:Frho_e4} can be expressed~as
\begin{align}
F_{\rho_e}(x) = 1 - \prod_{i=1}^{2} \left(1-F_{\rho_i}(x)\right) . 
\label{Eq:Frho_e6}
\end{align}
Next, by employing~\cite[Eq. (43)]{A:Analytical_Performance_Assessment_of_THz_Wireless_Systems} and after some algebraic manipulations, we can rewrite~\eqref{Eq:Frho_e6} as~\eqref{Eq:F_rho_e_final}. This concludes the~proof.  

\section*{Appendix B}

In the special case, in which either the S-R or the R-D link experience BM, according to~\eqref{Eq:Frho_e6}, the CDF of $\rho_e$ can be obtained~as
\begin{align}
F_{\rho_e}^{sc}=\left(1-F_{\rho_{i}}(x)\right) \left(1-F_{\rho_{j}^{wm}}(x)\right), 
\label{Eq:F_rho_e_sc_1}
\end{align}
where $i, j\in\{1, 2\}$ with $i\neq j$. Note that in~\eqref{Eq:F_rho_e_sc_1}, $F_{\rho_{i}}(x)$ denotes the CDF of the SNR of the link that suffers from BM, while $F_{\rho_{j}^{wm}}(x)$ is the one of the BM-free link. 

Next, we derive $F_{\rho_{j}}^{wm}(x)$~as
\begin{align}
F_{\rho_{j}^{wm}}(x) =  \Pr\left(\rho_{j}^{wm} \leq x\right),
\label{Eq:F_rho_wm}
\end{align}
where $\rho_{j}^{wm}$ is the SNR of the misalignment-free link, which can be expressed~as
\begin{align}
\rho_{j}^{wm} = \overline{\rho}_i \left|h_i^{(f)}\right|^2. 
\label{Eq:rho_wm}
\end{align}
From~\eqref{Eq:rho_wm},~\eqref{Eq:F_rho_wm} can be rewritten~as
\begin{align}
F_{\rho_{j}^{wm}}(x) =  \Pr\left(\left|h_i^{(f)}\right| \leq \sqrt{\frac{x}{\overline{\rho}_j}} \right),
\label{Eq:F_rho_j_wm_final}
\end{align}
or equivalently
\begin{align}
F_{\rho_{j}^{wm}}(x) = F_{\left|h_i^{(f)}\right|}\left(\sqrt{\frac{x}{\overline{\rho}_j}} \right).
\label{F_rho_j_wm_s2}
\end{align}
By taking into account~\eqref{Eq:F_hf1_definition},~\eqref{F_rho_j_wm_s2}, and \cite[Eq. (43)]{A:Analytical_Performance_Assessment_of_THz_Wireless_Systems}, we can express~\eqref{Eq:F_rho_e_sc_1} as~\eqref{Eq:F_rho_e_sc_final}. This concludes the~proof.

\section*{Appendix C} 

Following the same steps as Appendix A, the CDF of the end-to-end SNR in the absence of BM in both the S-R and R-D links can be obtained~as
$F_{\rho_{j}^{\text{id}}}(x) = 1 - \prod_{j=1}^{2}\left(1-F_{\rho_{j}^{wm}}(x)\right),$
which, by employing~\eqref{Eq:F_rho_j_wm_final}, we obtain~\eqref{Eq:F_rho_e_wm_final}. This concludes the~proof.

\balance
\bibliographystyle{IEEEtran}
\bibliography{IEEEabrv,References}
\balance

\end{document}